\documentclass[twocolumn]{aastex62}

\usepackage{amssymb}
\usepackage{amsmath}
\usepackage{hyperref}
\usepackage{natbib}
\usepackage{color}
\usepackage{graphicx}
\usepackage{amsmath}
\usepackage{bm}

\usepackage{bm}

\begin{document}

\title{The Radial Distribution of Ion-scale Waves in the Inner Heliosphere}

\correspondingauthor{Jinsong Zhao}
\email{js\_zhao@pmo.ac.cn}

\author{Wen Liu}
\affiliation{Key Laboratory of Planetary Sciences, Purple Mountain Observatory, Chinese Academy of Sciences, Nanjing 210023,  People's Republic of China}
\affil{School of Astronomy and Space Science, University of Science and Technology of China, Hefei 230026, People's Republic of China}

\author{Jinsong Zhao}
\affiliation{Key Laboratory of Planetary Sciences, Purple Mountain Observatory, Chinese Academy of Sciences, Nanjing 210023,  People's Republic of China}
\affil{School of Astronomy and Space Science, University of Science and Technology of China, Hefei 230026, People's Republic of China}

\author{Tieyan Wang}
\affiliation{School of Earth Science, Yunnan University, Kunming 650091, People's Republic of China}

\author{Xiangcheng Dong}
\affiliation{School of Earth Science, Yunnan University, Kunming 650091, People's Republic of China}
\affiliation{Key Laboratory of Planetary Sciences, Purple Mountain Observatory, Chinese Academy of Sciences, Nanjing 210023,  People's Republic of China}

\author{Justin C. Kasper}
\affiliation{ Climate and Space Sciences and Engineering, University of Michigan, Ann Arbor, MI 48109, USA}
\affiliation{ Smithsonian Astrophysical Observatory, Cambridge, MA 02138, USA}

\author{Stuart D. Bale}
\affiliation{Physics Department, University of California, Berkeley, CA 94720-7300, USA}
\affiliation{Space Sciences Laboratory, University of California, Berkeley, CA 94720-7450, USA}

\author{Chen Shi}
\affiliation{Key Laboratory of Planetary Sciences, Purple Mountain Observatory, Chinese Academy of Sciences, Nanjing 210023,  People's Republic of China}
\affil{School of Astronomy and Space Science, University of Science and Technology of China, Hefei 230026, People's Republic of China}

\author{Dejin Wu}
\affiliation{Key Laboratory of Planetary Sciences, Purple Mountain Observatory, Chinese Academy of Sciences, Nanjing 210023,  People's Republic of China}

\begin{abstract}

Determining the mechanism responsible for the plasma heating and particle acceleration is a fundamental problem in the study of the heliosphere. Due to efficient wave-particle interactions of ion-scale waves with charged particles, these waves are widely believed to be a major contributor to ion energization, and their contribution considerably depends on the wave occurrence rate. 
By analyzing the radial distribution of quasi-monochromatic ion-scale waves observed by the Parker Solar Probe, this work shows that the wave occurrence rate is significantly enhanced in the near-Sun solar wind, specifically 21\%$-$29\% below 0.3 au, in comparison to 6\%$-$14\% beyond 0.3 au. The radial decrease of the wave occurrence rate is not only induced by the sampling effect of a single spacecraft detection, but also by the physics relating to the wave excitation, such as the enhanced ion beam instability in the near-Sun solar wind. This work also shows that the wave normal angle $\theta$, the absolute value of ellipticity $\epsilon$, the wave frequency $f$ normalized by the proton cyclotron frequency $f_{\mathrm{cp}}$, and the wave amplitude $\delta B$ normalized by the local background magnetic field $B_0$ slightly vary with the radial distance. The median values of $\theta$, $|\epsilon|$, $f$, and $\delta B$  are about $9^\circ$, $0.73$, $3f_{\mathrm{cp}}$, and $0.01B_0$, respectively. 
Furthermore, this study proposes that the wave mode nature of the observed left-handed and right-handed polarized waves corresponds to the Alfv\'en ion cyclotron mode wave and the fast-magnetosonic whistler mode wave, respectively. 

\end{abstract}


\section{Introduction} \label{sec:introduction}

In weakly-collisional plasmas, one of the major mechanisms determining ion energization (ion plasma heating and ion acceleration) is wave-particle interactions of ion-scale waves \citep[e.g.,][]{2019LRSP...16....5V,2022ApJ...930...95Z}. Observations of highly anisotropic heavy ion temperatures with $T_\perp>T_\parallel$ from UVCS on SoHO \citep{1998ApJ...501L.127K} suggest ion-scale waves (in particular, ion cyclotron waves) are responsible for ion energization in the solar corona \citep[e.g.,][]{1999ApJ...518..937C,2001JGR...106.5649I,2001JGR...106.8357M,2001JGR...10629261L,2002JGRA..107.1147H}, where $T_\perp$ and $T_\parallel$ denote the temperatures perpendicular and parallel to the magnetic field, respectively. Also, ion cyclotron waves are usually proposed to locally energize and/or scatter solar wind ions \citep[e.g.,][]{2001JGR...106.8357M,2008PhRvL.101z1103K,2013PhRvL.110i1102K,2015ApJ...800L..31H,2021A&A...650A..10V,2022PhRvL.129p5101B,2022ApJ...926..185O,2022ApJ...924..112V}. These energizations are accomplished by cyclotron interactions between ions and ion-scale electromagnetic waves \citep[e.g.,][]{2002JGRA..107.1147H}. For example, from ion measurements by Helios and Wind, \cite{2001JGR...106.8357M} and \cite{2015ApJ...800L..31H} have shown observational evidences of pitch angle diffusion of solar wind ions induced by cyclotron wave-particle interactions. Recently, \cite{2022PhRvL.129p5101B} provided a similar pitch angle diffusion signature in the presence of ion cyclotron waves from Parker Solar Probe (PSP) observations, and they concluded that cyclotron heating occurs in the near-Sun solar wind.
Moreover, cyclotron resonant damping of ion-scale waves is proposed as one of the leading candidates for the dissipation of the solar wind turbulence \citep[e.g.,][]{1998JGR...103.4775L}.
Therefore, to better understand the micro wave-particle interaction process related to ion-scale waves in the solar wind, information on wave parameters needs to be explored.

Before PSP, ion-scale waves were detected at the heliocentric distances larger than 0.3 au \citep[e.g.][]{2009ApJ...701L.105J,2010JGRA..11512115J,2011ApJ...731...85H,2011ApJ...734...15P,2014ApJ...786..123J,2015JGRA..12010207B,2016ApJ...819....6W,2018JGRA..123.1715Z,2019ApJ...884L..53W}. For example, \cite{2009ApJ...701L.105J} performed a relatively comprehensive analysis for ion-scale waves in the solar wind near 1 au by using the STEREO observations, and they explored three main features of the observed waves \citep[also see][] {2014ApJ...786..123J}, i.e., the wave frequency locating around the proton cyclotron frequency, the waves propagating in the direction quasi-parallel and antiparallel to the background magnetic field, and the waves nearly behaving both left-handed (LH) and right-handed (RH) circularly polarized in the spacecraft frame. \cite{2010JGRA..11512115J} then observed similar LH and RH ion-scale waves in the solar wind near 0.3 au based on the MESSENGER measurements. These LH and RH waves were proposed to be ion cyclotron waves by \cite{2009ApJ...701L.105J,2010JGRA..11512115J}.
\cite{2015JGRA..12010207B} further studied the radial distribution of ion-scale waves from 0.3 to 0.7 au, and they found that these waves occupy about 6\% of the observation time of the MESSENGER. In addition to the polarization analysis relating to the power spectral matrix \citep{2009ApJ...701L.105J,2010JGRA..11512115J,2014ApJ...786..123J,2015JGRA..12010207B}, the analysis of the magnetic helicity signature also showed the prevalence of ion-scale waves in the solar wind \citep[e.g.,][]{2011ApJ...731...85H,2011ApJ...734...15P,2016ApJ...819....6W,2018JGRA..123.1715Z,2019ApJ...884L..53W}.

Recently, based on PSP measurements, ion-scale waves have been studied at the heliocentric distances below 0.3 au \citep{2019Natur.576..237B,2020ApJ...899...74B,2020ApJS..246...66B,2022PhRvL.129p5101B,2020ApJS..248....5V,2022ApJ...924..112V,2020ApJ...897L...3H,2021ApJ...915L...8D,2021ApJ...909....7K,2021ApJ...908L..19S,2021A&A...650A..10V}. Through statistical analysis of the data in the first Encounter of PSP, \cite{2020ApJS..246...66B} explored an unexpected high occurrence rate of ion-scale waves, for example, waves can occupy 30\%-50\% of the time in radial magnetic field intervals. \cite{2020ApJS..248....5V} analyzed two ion-scale wave events containing proton beams, and they showed that the observed LH and RH ion-scale waves correspond to the ion cyclotron and fast-magnetosonic waves, respectively, which are both produced by proton beam instabilities. \cite{2020ApJ...899...74B}, \cite{2021ApJ...909....7K}, and \cite{2021ApJ...908L..19S} also showed a close connection of the observed LH and RH ion-scale waves with the ion cyclotron wave and fast-magnetosonic wave. Additionally, \cite{2020ApJ...897L...3H} and \cite{2021ApJ...915L...8D} provided observational evidence of coexistence of quasi-parallel and quasi-perpendicular ion-scale waves.

In this paper, following recent observations of ion-scale waves by PSP \citep[e.g.,][]{2020ApJ...899...74B,2020ApJS..246...66B,2020ApJS..248....5V}, we aim to explore the radial distribution of ion-scale waves in the inner heliosphere based on five years of PSP observations. PSP provides a unique opportunity to detect the ion-scale wave that extends from the solar corona to the solar wind near the Earth \citep{2021PhRvL.127y5101K}.

This paper is organized as follows. Section 2 introduces the data and methodology. Section 3 shows the radial distributions of the observed waves. Section 4 identifies the wave mode nature. The discussion and summary are given in Section 5 and 6, respectively.

\section{Data and Methodology}
Since this paper pays attention to quasi-monochromatic ion-scale waves and the related background plasma environment, we use both magnetic field and plasma parameters measured by the FIELDS instrument \citep[]{2016SSRv..204...49B} and Solar Wind Electrons Alphas and Protons (SWEAP) instrument \citep[]{2016SSRv..204..131K}. We consider data during Encounters 1-11, when the FIELDS magnetometer provide high-resolution data with sample rate ($\sim 73-293$ Hz) higher than the frequency ($\sim 0.1-10$ Hz) of ion-scale waves of interest. We use data from both the Solar Probe Cup \citep[SPC;][]{2020ApJS..246...43C} and the Solar Probe ANalyzers for Ions \citep[SPAN-I;][]{2022ApJ...938..138L} to collect the solar wind speed, number density, and temperature. Through SPC and SPAN-I measurements, we can cross check the plasma parameters. 

In order to identify quasi-monochromatic ion-scale waves from PSP observations \citep[e.g.,][]{2020ApJS..246...66B,2020ApJS..248....5V}, the wave analysis is performed by the method similar to \cite{2019GeoRL..46.4545Z,2020ApJ...890...17Z} and \cite{2021ApJ...908L..19S}, and then ion-scale wave events are picked under the following identification procedures.

The first procedure is to obtain the magnetic fluctuations $\delta \bm {B}(f,t)$ with varying frequency $f$ and time $t$. The time series magnetic field data are divided into consecutive overlapping time windows. The sliding time window $t_{\mathrm{sliding}}$ is 35 minutes and the overlapping time is 5 minutes. 
In each time window, the background magnetic field $\bm {B_0}$ is given by smoothly averaging $\bm {B}$ at the time of the maximum of 20 s and $20/f_{\mathrm{cp}}$, where $f_{\mathrm{cp}}$ is the proton cyclotron frequency. According to the frequencies of observed ion-scale waves with $f \gtrsim 0.5 f_{\mathrm{cp}}$ in the solar wind between 0.3 and 0.7 au \citep{2010JGRA..11512115J,2015JGRA..12010207B} and $f$ being of the order of 1 Hz in the solar wind below 0.3 au \citep[e.g.,][]{2020ApJS..246...66B}, this averaging time is about at least 10 times longer than the wave period. 
Then, the wavelet transform is used to deal with $\bm {B}(t)$ during $t_{\mathrm{sliding}}$ via \citep{1998BAMS...79...61T}

\begin{equation}
 {\bm W} (s,t) = \sum_{i=0}^{N-1 }
 \psi \left( \frac{t_i-\tau}{s}  \right) {\bm B} \left( t_i \right),
\end{equation}
where $s$ denotes the wavelet scale, and $\tau$ denotes a time parameter. The center part with 30 minutes in $t_{\mathrm{sliding}}$ is chosen to perform the wave analysis (the two edge parts with 2.5 minutes are not used to analyze the waves due to the edge effects in the wavelet transform). Also, the wavelet transform is dealt with in the magnetic field-aligned coordinates defined as $( {\bm e_1} \equiv ({\bm e_{B_0}}\times {\bm e_R})\times {\bm e_R}, {\bm e_2} \equiv {\bm e_{B_0}}\times {\bm e_R},  {\bm e_3} \equiv {\bm e_{B_0}})$, where ${\bm e_{B_0}}\equiv \bm {B_0}/|\bm {B_0}|$ is the unit vector of the background magnetic field, and ${\bm e_R}={\bm R}/|{\bm R}|$ is the unit radial vector. Note that the Morlet wavelet $\psi (\eta) = \pi^{-1/4} e^{i\omega_0 \eta} e^{-\eta^2/2}$  is implemented to perform the transform, where $\omega_0$ represents the nondimensional frequency and $\omega_0=6$ is used \citep{1998BAMS...79...61T}. 
Hence, $\delta \bm {B}(f,t) = {\bm W} (s,t)$ can be obtained through the relation of $f=(\omega_0 + \sqrt{\omega_0^2+2})/(4\pi s)$ \citep{1998BAMS...79...61T}.
For giving sufficient information of the wave frequency distribution, the wavelets with the number being the nearest large integer of $36\times \mathrm{log}_{10}(f_{\mathrm{max}}/f_{\mathrm{min}})$ are used to equally separate the frequency region between $f_{\mathrm{min}}=0.1 f_{\mathrm{cp}}$ and $f_{\mathrm{max}} = \mathrm{max} \left(20~\mathrm{Hz},10 f_{\mathrm{cp}} \right)$ in the logarithmic space. 
Moreover, when $f_{B}<20 f_{\mathrm{cp}}$ ($f_{B}$ is the sample frequency of $\bm {B}$), $f_{\mathrm{max}}$ is fixed at $30$ Hz that can well capture all quasi-monochromatic ion-scale waves.

The second procedure is to give the wave parameters by using the singular value decomposition (SVD) method. \cite{2003RaSc...38.1010S} have shown that using the SVD for the matrix in the following equation

\begin{equation}
\left(
\begin{array}{ccc}
\mathfrak{R}S_{\mathrm{11}} & 
\mathfrak{R}S_{\mathrm{12}} & 
\mathfrak{R}S_{\mathrm{13}} \\
\mathfrak{R}S_{\mathrm{21}} & 
\mathfrak{R}S_{\mathrm{22}} & 
\mathfrak{R}S_{\mathrm{23}} \\
\mathfrak{R}S_{\mathrm{31}} & 
\mathfrak{R}S_{\mathrm{32}} & 
\mathfrak{R}S_{\mathrm{33}} \\
0 & 
-\mathfrak{I}S_{\mathrm{12}} & 
-\mathfrak{I}S_{\mathrm{13}} \\
-\mathfrak{I}S_{\mathrm{21}} & 
0 & 
-\mathfrak{I}S_{\mathrm{23}} \\
-\mathfrak{I}S_{\mathrm{31}} & 
-\mathfrak{I}S_{\mathrm{32}} & 
0 \\
\end{array}
\right) \cdot {\bm k} =0,
\label{eigen_equation}
\end{equation}
one can obtain three eigenvalues (i.e., the largest eigenvalue $\lambda_l$, the intermediate eigenvalue $\lambda_m$, and the smallest eigenvalue $\lambda_s$) and the corresponding unit vector ${\bm \kappa}={\bm k}/k$ associating with $\lambda_s$. Here, $S_{\mathrm{ij}} = \langle \delta B_i\delta B^*_j \rangle$ (the superscript ``*'' denotes the complex conjugate, and the symbol ``$\langle \rangle$'' denotes an average over 8 periods, i.e., $1/f$), and $\mathfrak{R}$ and $\mathfrak{I}$ represent the real and imaginary parts of the variable $S_{\mathrm{ij}}$, respectively. Then, it can directly give several key parameters that describe the fluctuations, i.e., the degree of polarization $\mathrm{DOP}(f,t)$, the wave normal angle $\theta(f,t)$, and the ellipticity $\epsilon(f,t)$, through the following expressions \citep[e.g.,][]{1980GeoJ...61..115S,2003RaSc...38.1010S},
\begin{eqnarray}
\mathrm{DOP} &=& 
\sqrt{ \frac{3}{2} \frac { \mathrm{trace}({\bm S}^2)} {\mathrm{trace}({\bm S})^2} -\frac{1}{2}}
=
\frac{ \sqrt{ \sum_{j,k} \left(\lambda_j - \lambda_k \right)^2 }} 
{ 2 \left( \sum_{j}\lambda_j \right)},
\\
\theta &=& \mathrm{arctan} \left( \frac{\sqrt{\kappa_1^2 + \kappa_2^2}}{\kappa_3} \right),
\\
\epsilon &=& \frac{\lambda_m}{\lambda_l},
\end{eqnarray}
where the subscripts ``$j$'' and ``$k$'' correspond to ``$l$'', ``$m$'', or ``$s$'', and $\kappa_{1}$, $\kappa_{2}$, and $\kappa_{3}$ denote the three components of ${\bm \kappa}$ in the (${\bm e_1}$, ${\bm e_2}$, ${\bm e_3}$) coordinate system.
For describing left-hand or right-hand polarization for the waves, $\epsilon$ is further given as $\epsilon =(\lambda_m/\lambda_l)\times \mathrm{ sign} (\mathfrak{I}S_{12})$, which can explore the magnetic polarization of quasi-parallel/antiparallel waves. Thus, we can conveniently identify the wave event by combining these parameters. For example, for quasi-monochromatic and quasi-parallel ion-scale waves, they have high $\mathrm{DOP}$, small $\theta$, and large $|\epsilon|$ (the quantitative criteria for these parameters are proposed in the next procedure). However, we cannot distinguish the wave propagating direction parallel or antiparallel to the background magnetic field through SVD method  \citep[e.g.,][]{2003RaSc...38.1010S}. Note that DOP determines coherence of the fluctuations. 

The third procedure is to collect wave and plasma parameters:

(1) For collecting wave parameters, $\mathrm{DOP}(f,t)$, $\theta(f,t)$, and $\epsilon(f,t)$ are averaged over the time interval $\Delta t_{\mathrm{coll}}=10$ s (this time is chosen by the experience), and $\mathrm{DOP}(f)$, $\theta(f)$, and $\epsilon(f)$ at each $f$ are obtained. We note that $\Delta t_{\mathrm{coll}}$ is approximately half of the mean duration of the wave events ($\sim 21$ seconds) found by \cite{2020ApJS..246...66B}, who performed the statistics of ion-scale waves in Encounter 1.

(2) The wave is selected based on the criteria of $\mathrm{DOP}(f) > 0.75$, $|\epsilon(f)| > 0.60$, and $\theta(f)<30^\circ$. These criteria differ somewhat from those used in the statistical work of \cite{2015JGRA..12010207B}, where $\mathrm{DOP}(f) > 0.7$, $|\epsilon(f)| > 0.65$, and $\theta(f)<40^\circ$. However, they can effectively select quasi-monochromatic and quasi-parallel ion-scale waves of interest (see discussions in Section 5). Three other criteria are further used to pick the event. 
The first extra criterion is $T\geq 20$ s. 
Here, $T$ is defined as the total time of continue time intervals $\Delta t_{\mathrm{coll}}$ when the criteria of $\mathrm{DOP}$, $|\epsilon(f)|$, and $\theta$ are satisfied, and $T=n_{\mathrm{\Delta t}} \Delta t_{\mathrm{coll}}$ for  the number of continue $\Delta t_{\mathrm{coll}}$ being $n_{\mathrm{\Delta t}}$. The $T\geq 20$ s criterion indicates that one whole wave event survive at least two successive $\Delta t_{\mathrm{coll}}$. 
The second extra criterion is that the wave event must cover at least $n_{f}=4$ wavelets during $\Delta t_{\mathrm{coll}}$ ($n_{f}$ denotes the number of the continue wavelets in the presence of the waves of interest, and the choice of $n_{f}=4$ is based on the experience). 
The third extra criterion is that when there exist multi waves with the same sign of $\epsilon$ (i.e., $\epsilon>0$ or $\epsilon<0$) but with different frequency bands during $\Delta t_{\mathrm{coll}}$, only the waves covering the maximum number of wavelets are collected. This criterion can roughly pick the waves with strongest amplitude. In fact, comparing to one band wave, multi band waves with the same polarization are uncommon. Under these criteria, we finally obtain the information of $\mathrm{DOP}$, $\epsilon$, $\theta$, and the wave frequency band $\Delta f = f_{\mathrm{upper}} - f_{\mathrm{lower}}$, where $f_{\mathrm{upper}}$ and $f_{\mathrm{lower}}$ denote the upper and lower boundaries of the wave frequency band.
Using the identified wave frequency band (i.e.,  $ f_{\mathrm{lower}} - f_{\mathrm{upper}}$), we filter the magnetic fields $\bm {B}$ and average the peaks of filtered magnetic field fluctuations as the proxy of the wave amplitude $\delta B$.

(3) Plasma parameters (including the solar wind speed, number density, and temperature) are directly collected by averaging over an extended time interval of 50 seconds (the choice of this time interval is based on experience). In the absence of valid plasma measurements during this time interval, we set plasma parameters as null values.

Through aforementioned procedures to process the high-resolution magnetic field data from the FIELDS flux-gate magnetometer \citep[]{2016SSRv..204...49B}, we obtain a wave dataset containing 409,888 data (the duration of each data is $\Delta t_{\mathrm{coll}}$) for ion-scale waves of interest during Encounters 1-11. The total duration of the waves is approximately 47.4 days, occupying $\sim 20\%$ of the total observation time of PSP.

\section{Radial distributions of observed ion-scale waves}

\begin{figure}
\includegraphics[width=8.5 cm]{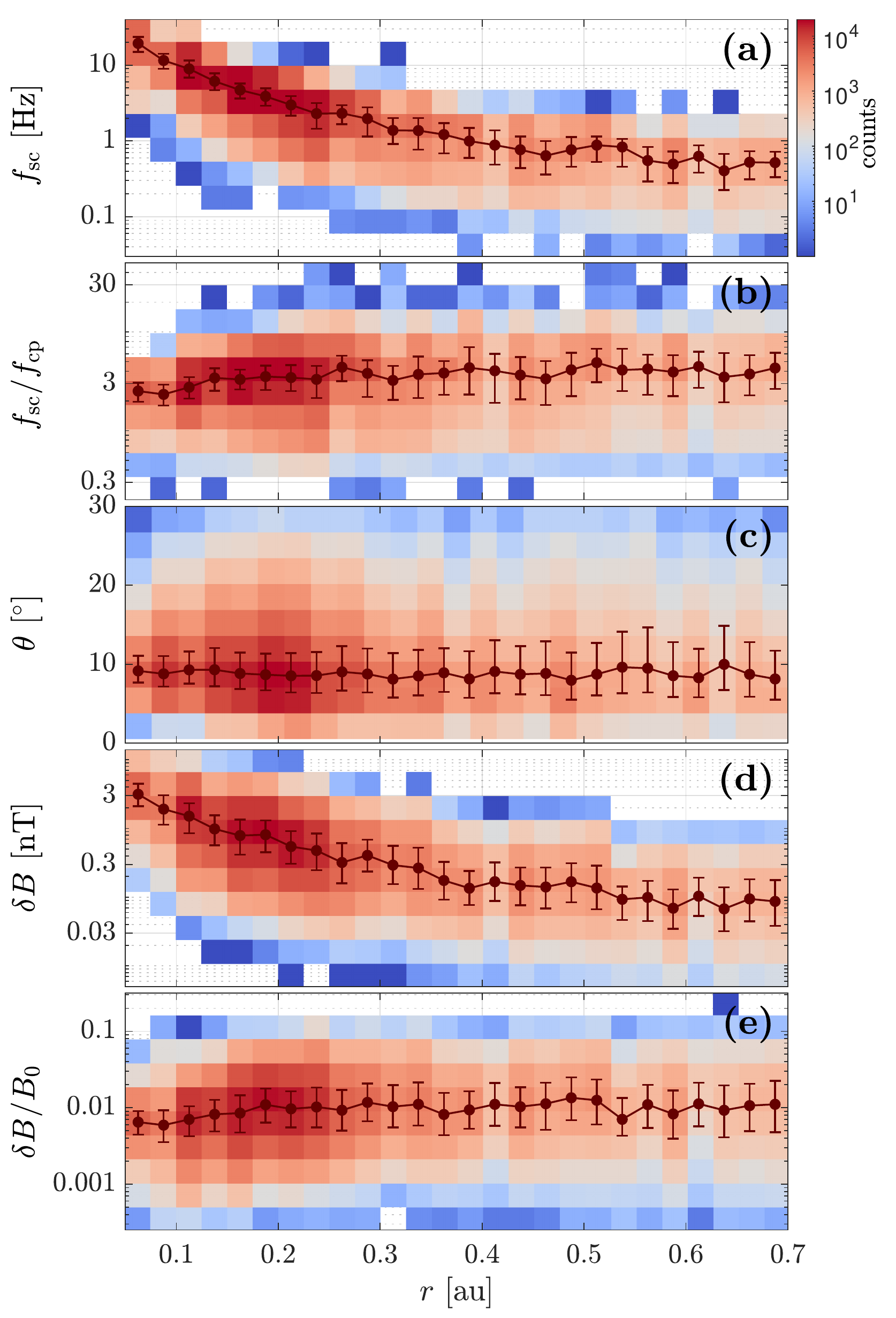}
\caption{
Radial distributions of ion-scale waves: (a) the weighted frequency, $f_{\mathrm{sc}}$; (b) $f_{\mathrm{sc}}$ normalized by the proton cyclotron frequency, $f_{\mathrm{sc}}/f_{\mathrm{cp}}$; (c) the weighted wave normal angle, $\theta$; (d) the wave amplitude, $\delta B$; and (e) $\delta B$ normalized by the background magnetic field, $\delta B/B_0$. The filled points represent the median of each variable (the second quartile) in each radial bin, and the lower/upper error bar denotes the difference between the lower/upper quartile and the  second quartile. 
\label{fig:radial_distribution}}
\end{figure}

\begin{figure}
\includegraphics[width=8.5 cm]{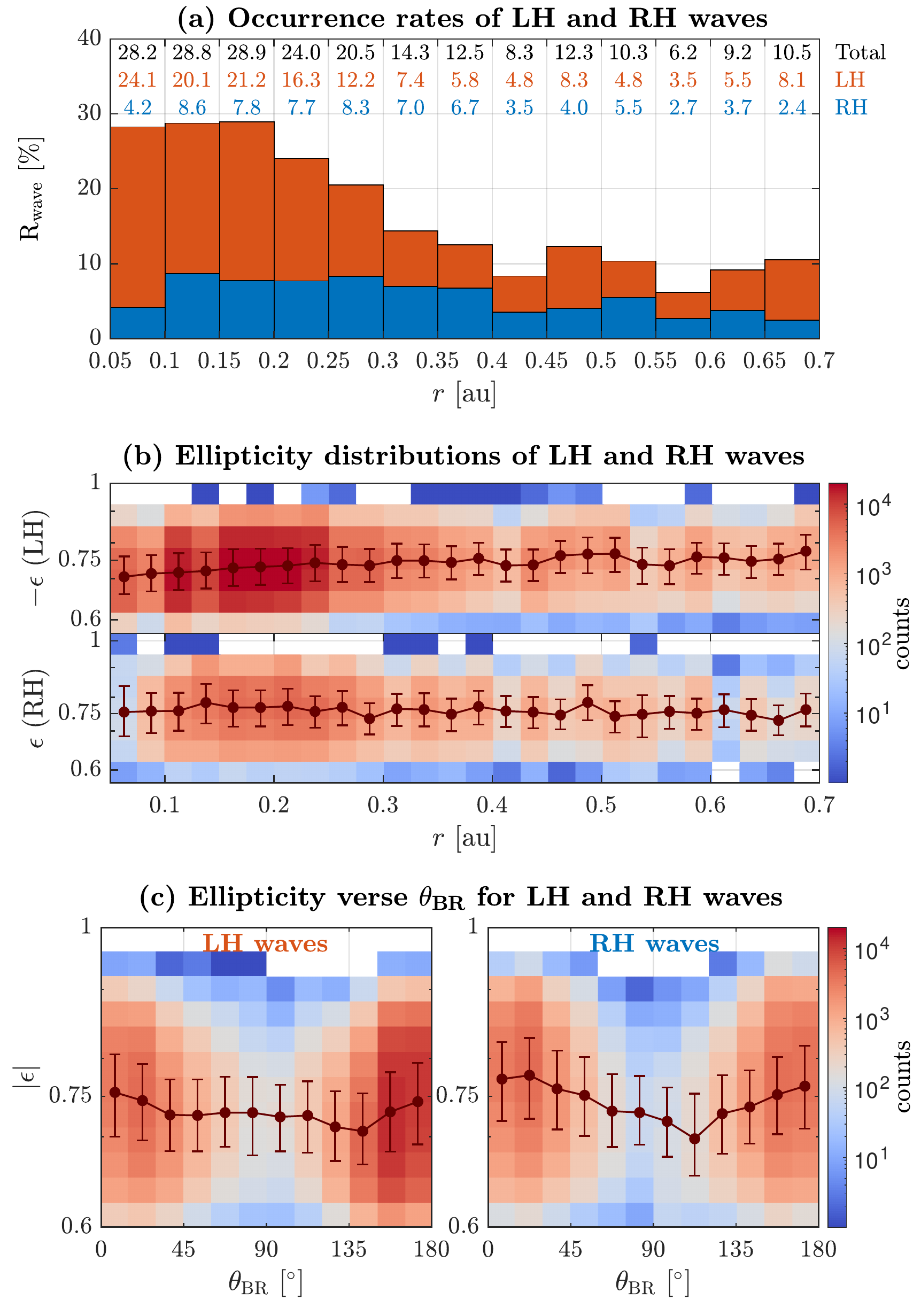}
\caption{
(a) Occurrence rates of LH (red color) and RH (blue color) polarized waves. (b) The radial distributions of ellipticity $\epsilon$ of LH and RH waves. (c) Joint distributions of ellipticity $|\epsilon|$ and angle $\theta_{\mathrm{BR}}$ between the magnetic field ${\bm B}$ and the radial vector ${\bm R}$ for LH and RH waves. The descriptions of the filled points and error bars in Panels (b) and (c) are the same as those in Figure \ref{fig:radial_distribution}.
\label{fig:LH_RH}}
\end{figure}

\begin{figure*}
\includegraphics[width=17.5 cm]{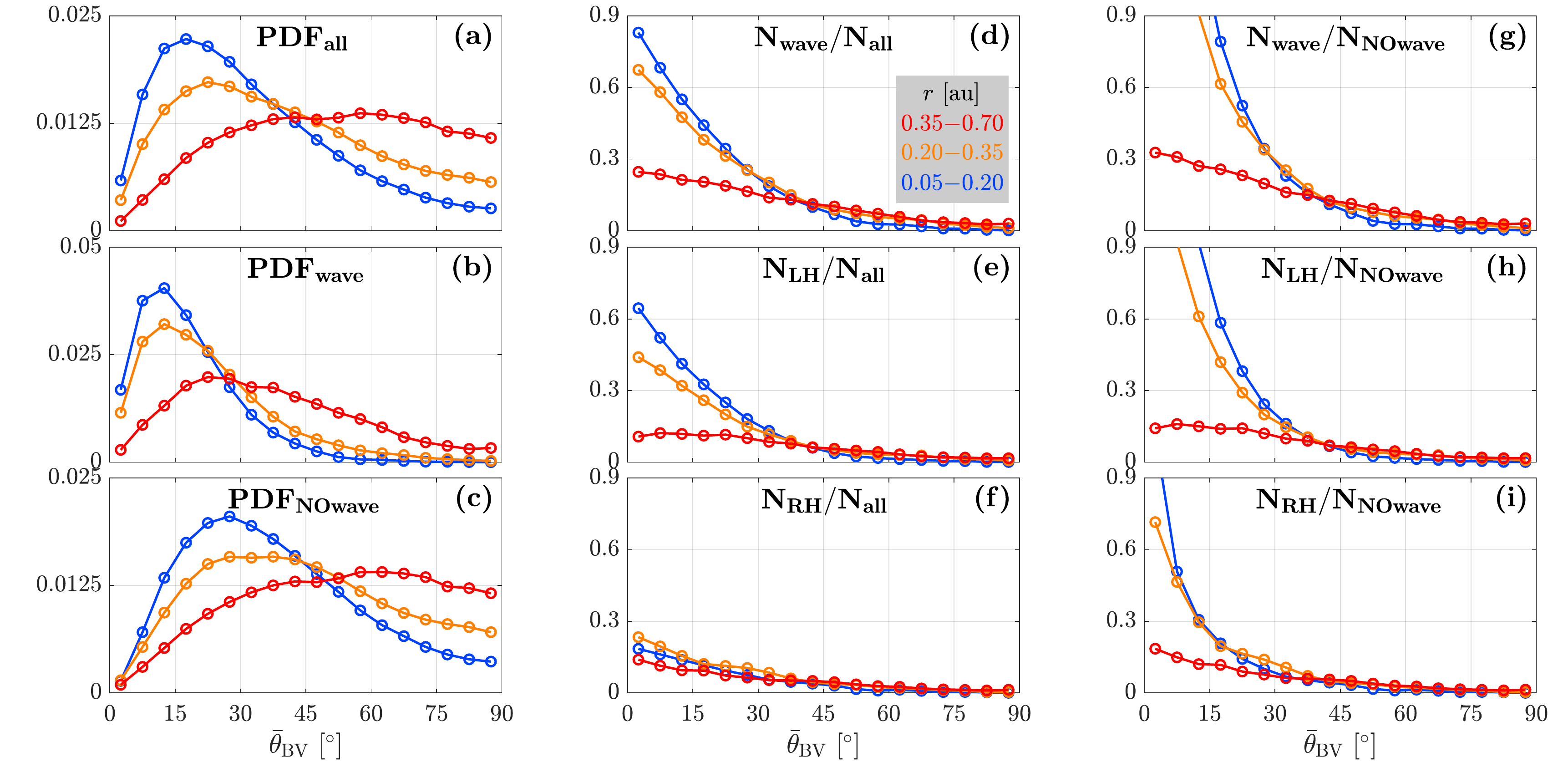}
\caption{
The PDF distributions of (a) all data $\mathrm{PDF_{all}}({\bar \theta}_{\mathrm{BV}},r)$, (b) the wave data $\mathrm{PDF_{wave}}({\bar \theta}_{\mathrm{BV}},r)$, and (c) the data without the waves $\mathrm{PDF_{NOwave}}({\bar \theta}_{\mathrm{BV}},r)$ in three radial regimes: $r=0.05-0.2$ au (blue), $r=0.2-0.35$ au (orange), and $r=0.35-0.7$ au (red). 
The distributions of the occurrence rate $N_{\mathrm{wave}}({\bar \theta}_{\mathrm{BV}},r)/N_{\mathrm{all}}({\bar \theta}_{\mathrm{BV}},r)$ of all waves (d), $N_{\mathrm{LH}}({\bar \theta}_{\mathrm{BV}},r)/N_{\mathrm{all}}({\bar \theta}_{\mathrm{BV}},r)$ of LH waves (e), and $N_{\mathrm{RH}}({\bar \theta}_{\mathrm{BV}},r)/N_{\mathrm{all}}({\bar \theta}_{\mathrm{BV}},r)$ of RH waves (f).
Panels (g)-(i): similar to Panels (d)-(f), but for the ratio between the data numbers with and without the waves. ${\bar \theta}_{\mathrm{BV}}=0^\circ-90^\circ$ is equally separated into 18 bins with the width of $5^\circ$. The solar wind velocity ${\bf V_0}$ used here comes from SPAN-I measurements.
\label{fig:sampling}}
\end{figure*}

The radial distributions of observed ion-scale waves are shown in Figure \ref{fig:radial_distribution}. The data is redistributed into 26 radial bins in the heliocentric distance $r$ between $0.05$ and  $0.7$ au, and each bin occupies equally 0.025 au. The median value (also known as second quartile) of wave parameters in each bin is denoted by the dark red point, and the lower (upper) error bar is the difference between the second quartile and the lower (upper) quartile, where the lower (upper) quartile corresponds to the median of the set of values less (greater) than the second quartile.

Figure \ref{fig:radial_distribution}(a) shows the radial distribution of the weighted wave frequency $f_{\mathrm{sc}}$ in the spacecraft frame. Here,  $f_{\mathrm{sc}}$ is defined as 
$f_{\mathrm{sc}} = \sum_{f=f_{\mathrm{lower}}}^{f_{\mathrm{upper}}} \left[f  P_{\mathrm{B_\perp}}(f) \right]$ $/ \sum_{f=f_{\mathrm{lower}}}^{f_{\mathrm{upper}}}  P_{\mathrm{B_\perp}}(f)$,
where $P_{\mathrm{B_\perp}} (f) = 2\delta t [\delta B_1^2(f) + \delta B_2^2(f)]$ is the power spectral density of the perpendicular magnetic field fluctuation \citep[e.g.,][]{2016ApJ...824...47L}, $\delta B_{1,2}(f)$ is the average value at each $f$ in the wave frequency band during $\Delta t_{\mathrm{coll}}$, and $\delta t$ is the time resolution of the magnetic field used in the wavelet transform. 
In fact, $2\delta t$ in the expression of $P_{\mathrm{B_\perp}} (f) $ does not contribute to the calculation of $f_{\mathrm{sc}}$. From Figure \ref{fig:radial_distribution}(a), we see that $f_{\mathrm{sc}}$ is considerably decreasing with increasing $r$ as $r\lesssim 0.4$ au, and it slightly varies at larger $r$ (e.g., $r \gtrsim 0.5$ au). 
Actually, the observed frequency $f_{\mathrm{sc}}$ consists of the wave frequency $f_{\mathrm{pl}}$ in the plasma frame and the Doppler shift frequency ${\bm V_0}\cdot {\bm k}/(2\pi)$ resulting from the solar wind flow ${\bm V_0}$, that is, $f_{\mathrm{sc}} =  f_{\mathrm{pl}} + {\bm V_0}\cdot {\bm k}/(2\pi)$. 
When the solar wind is super-Alfv\'enic ($|{\bm V_0}|$ is much larger than the local Alfv\'en speed $V_A$), the Doppler shift frequency is the dominant part in $f_{\mathrm{sc}}$. We will remove the Doppler shift frequency and estimate $f_{\mathrm{pl}}$ in Section 4.

Figure \ref{fig:radial_distribution}(b) exhibits $f_{\mathrm{sc}}$ normalized by the proton cyclotron frequency $f_{\mathrm{cp}}=eB_0/(2\pi m_p)$. This distribution is slightly varying with $r$, and the median value of $f_{\mathrm{sc}}/f_{\mathrm{cp}}$ is around $3$ throughout radial distances. 

Figure \ref{fig:radial_distribution}(c) presents the radial distribution of the weighted wave normal angle $\theta$, which is defined as
$\theta =  \sum_{f=f_{\mathrm{lower}}}^{f_{\mathrm{upper}}}  \left[\theta(f) P_{\mathrm{B_\perp}}(f) \right] /  \sum_{f=f_{\mathrm{lower}}}^{f_{\mathrm{upper}}}  P_{\mathrm{B_\perp}} (f)$.
This distribution is nearly unchanged with varying $r$, and the median value is about $9^\circ$. 
This characteristic value ($\theta\simeq 9^\circ$) is nearly consistent with previous statistical results explored by \cite{2015JGRA..12010207B} and \cite{2020ApJS..246...66B}. For example, \cite{2015JGRA..12010207B} showed the peak of the distribution of $\theta$ residing $\sim 7^\circ$, and \cite{2020ApJS..246...66B} exhibited the median value in their statistical $\theta$ in Encounter 1 being $\sim 5^\circ$.

Figure \ref{fig:radial_distribution}(d) shows the radial distribution of the wave amplitude $\delta B$. $\delta B$ obviously decreases with increasing $r$ in the near-Sun solar wind with $r\lesssim 0.4$ au. For example, the median $\delta B$  is about 3 nT at $r \sim 0.075$ au and 0.3 nT at $r \sim 0.4$ au. At $r \gtrsim 0.4$ au, the wave amplitude slightly decreases with increasing $r$. 

Figure \ref{fig:radial_distribution}(e) shows the radial distribution of the wave amplitude $\delta B$ normalized by $B_0$. This figure exhibits that $\delta B/B_0$ is slightly varying with $r$, and the median value is around $0.01$.

Figure \ref{fig:LH_RH} further classifies observed ion-scale waves into LH and RH polarized waves and exhibit their statistical distributions. The LH and RH waves correspond to the waves with $\epsilon<-0.6$ and $\epsilon>0.6$, respectively.

Figure \ref{fig:LH_RH}(a) presents the radial distribution of occurrence rates $R_{\mathrm{LH}}$ and $R_{\mathrm{RH}}$ (defined as the ratio between wave occurrence time and total observation time in each bin) of LH and RH waves. Both $R_{\mathrm{LH}}$ and $R_{\mathrm{RH}}$ enhance at $r \lesssim 0.3$ au, in which $R_{\mathrm{LH}}+R_{\mathrm{RH}}$ is about $21\%-29\%$ ($R_{\mathrm{LH}}\sim 12\%-24\%$ and $R_{\mathrm{RH}}\sim 4\%-9\%$) in the defined bins, comparing $R_{\mathrm{LH}}+R_{\mathrm{RH}} \sim 6\%-14\%$ in the defined bins at $r \gtrsim 0.3$ au. The total wave occurrence rate below 0.3 au (beyond 0.3 au) is about 26\% (11\%). Our wave occurrence rate beyond 0.3 au is larger than that explored by \cite{2015JGRA..12010207B}, who found the occurrence rate being around $6\%$ between 0.3 and 0.7 au by using nearly 4 years data from MESSENGER mission. The reason of this difference may come from different wave identification methods and different datasets used in this work and the work of \cite{2015JGRA..12010207B}. 
When our data are limited to the same time interval as \cite{2020ApJS..246...66B}, the wave occurrence rate approximates 30\%, and it is nearly 43\% at $|\theta_{\mathrm{BV}} - 90^\circ| > 60^\circ$ ($\theta_{\mathrm{BV}}$ is the angle between ${\bm B_0}$ and ${\bm V_0}$), which supports the finding of high occurrence rate of ion-scale waves by \cite{2020ApJS..246...66B}.

Figure \ref{fig:LH_RH}(b) presents the radial distributions of ellipticity $\epsilon$ of LH and RH waves. Both LH and RH waves have similar ellipticity distributions that are nearly unchanged with varying radial distance. The median value of ellipticity $\epsilon_{m}$ is around $-0.73$ for LH waves and $ 0.76$ for RH waves.

Figure \ref{fig:LH_RH}(c) presents joint distributions of $|\epsilon|$ and $\theta_{\mathrm{BR}}$ (the angle between ${\bm B_0}$ and ${\bm R}$) for both LH and RH waves. This figure shows that the majority of both LH and RH waves concentrate in the near-radial magnetic field environments where $\theta_{\mathrm{BR}}<30^\circ$ and $\theta_{\mathrm{BR}}>150^\circ$. The similar distributions have been reported by previous studies \citep[e.g.,][]{2010JGRA..11512115J,2015JGRA..12010207B,2020ApJS..246...66B}.

The most important finding explored by Figure \ref{fig:LH_RH} is the enhancement of the wave occurrence rate in the near-Sun solar wind. However, because single-point spacecraft measurements provide only a reduced spectrum, the strength of the spectrum of quasi-parallel/antiparallel waves decreases when the deviation between ${\bm B_0}$ and ${\bm V_0}$ increases, while the strength of the spectrum of the background anisotropic turbulence increases  \citep{2020ApJS..246...66B}. As a result, the sampling effect of quasi-parallel and antiparallel ion-scale waves due to single spacecraft measurements can lead to the wave signature being masked by the background anisotropic turbulence \citep{2020ApJS..246...66B}, especially in the solar wind environment where there is a large deviation of ${\bm B_0}$ from ${\bm V_0}$. Normally, the deviation between ${\bm B_0}$ from ${\bm V_0}$ enhances with increasing $r$ due to the configuration of the solar wind magnetic field (i.e., the Parker spiral). The sampling effect may be one reason for forming the radial distribution of the wave occurrence rate shown in Figure \ref{fig:LH_RH}. Therefore, we proceed to analyze the dependence between the occurrence rate and the sampling effect.

Considering two facts, i.e., (1) the key factor being the deviation of ${\bm B_0}$ from ${\bm V_0}$, and (2) the radial change of the wave occurrence rate, the data are cut into the bins with different $r$ and different ${\bar \theta}_{\mathrm{BV}}$. Here, ${\bar \theta}_{\mathrm{BV}}$ is defined as ${\bar \theta}_{\mathrm{BV}}=\theta_{\mathrm{BV}}$ at $\theta_{\mathrm{BV}}\leq 90^\circ$ and ${\bar \theta}_{\mathrm{BV}}=180^\circ - \theta_{\mathrm{BV}}$ at $\theta_{\mathrm{BV}}> 90^\circ$ to represent the deviation of ${\bm B_0}$ from ${\bm V_0}$. In order to have sufficient number of the data in different $r$ regimes, we separate the radial distance between $0.05$ and $0.7$ au into three regimes, i.e., $r=0.05-0.2$ au (including 196,486  wave data of interest and 682,398 data of total PSP observations), $0.2-0.35$ au (including 147,641 wave data and 707,759 total PSP data), and $0.4-0.7$ au (including 62,841 wave data and 646,560 total PSP data). We also divide ${\bar \theta}_{\mathrm{BV}}$ into 18 equal regimes with $(j-1) \times 5^\circ < {\bar \theta}_{\mathrm{BV}} <j \times 5^\circ$ ($j=1, ...,18$). The number of the total PSP data, $N_{\mathrm{all}}({\bar \theta}_{\mathrm{BV}},r)$, exceeds 10,000 in most of the bins, and 4 bins contain $N_{\mathrm{all}}({\bar \theta}_{\mathrm{BV}},r)$ roughly between 3000 and 10,000. The number of the wave data $N_{\mathrm{wave}}({\bar \theta}_{\mathrm{BV}},r)$ in our defined bins distributes from 11 to 39,792.


For exhibiting the distribution feature of $N_{\mathrm{all}}({\bar \theta}_{\mathrm{BV}},r)$ and $N_{\mathrm{wave}}({\bar \theta}_{\mathrm{BV}},r)$, Figures \ref{fig:sampling}(a) and (b) present the probability distribution functions (PDFs) defined by
\begin{eqnarray}
\mathrm{PDF}_{i}({\bar \theta}_{\mathrm{BV}},r) &=& 
\frac{N_{i}({\bar \theta}_{\mathrm{BV}},r)}
        { \Delta {\bar \theta}_{\mathrm{BV}}
        \sum_{{\bar \theta}_{\mathrm{BV}}} N_{i}({\bar \theta}_{\mathrm{BV}},r) }, 
\end{eqnarray}
for both the total PSP data ``$i =\mathrm{all}$'' and the wave data ``$i =\mathrm{wave}$''. Figure \ref{fig:sampling}(c) further presents the PDF of the number of the data without the waves in the given bins, in which  $N_{\mathrm{NOwave}}({\bar \theta}_{\mathrm{BV}},r)=N_{\mathrm{all}}({\bar \theta}_{\mathrm{BV}},r)-N_{\mathrm{wave}}({\bar \theta}_{\mathrm{BV}},r)$.


Figure \ref{fig:sampling}(a) shows that $\mathrm{PDF_{all}}$ concentration is around ${\bar \theta}_{\mathrm{BV}}\sim 17.5^\circ$ in the $r$ regime close to the Sun, and the peak of $\mathrm{PDF_{all}}$ shifts to large ${\bar \theta}_{\mathrm{BV}}$ ($ \sim 60^\circ$) in the $r$ regime far away from the Sun. Figure \ref{fig:sampling}(b) shows that the peaks of $\mathrm{PDF_{wave}}$ in $r=0.05-0.2$ au and $0.2-0.35$ au regimes are around ${\bar \theta}_{\mathrm{BV}}\sim 12.5^\circ$, 
while the corresponding peak in $r=0.35-0.7$ au regime is around ${\bar \theta}_{\mathrm{BV}}\sim 22.5^\circ$.
$\mathrm{PDF_{NOwave}}$ shown in Figure \ref{fig:sampling}(c) is more similar to $\mathrm{PDF_{all}}$ than $\mathrm{PDF_{wave}}$, and the reason is that $\mathrm{PDF_{NOwave}} \sim \mathrm{PDF_{all}}$ in most of  ${\bar \theta}$ bins.


To explore the sampling effect, Figure \ref{fig:sampling}(d) presents the occurrence rate of the waves, $N_{\mathrm{wave}}({\bar \theta}_{\mathrm{BV}},r)$ $/N_{\mathrm{all}}({\bar \theta}_{\mathrm{BV}},r)$, as a function of ${\bar \theta}_{\mathrm{BV}}$ in the three radial regimes. Figure \ref{fig:sampling}(e) and (f) present the occurrence rates of LH and RH waves,  $N_{\mathrm{LH}}({\bar \theta}_{\mathrm{BV}},r)$ $/N_{\mathrm{all}}({\bar \theta}_{\mathrm{BV}},r)$ and $N_{\mathrm{RH}}({\bar \theta}_{\mathrm{BV}},r)$ $/N_{\mathrm{all}}({\bar \theta}_{\mathrm{BV}},r)$, as a function of ${\bar \theta}_{\mathrm{BV}}$, respectively, where $N_{\mathrm{LH}}({\bar \theta}_{\mathrm{BV}},r)$ and $N_{\mathrm{RH}}({\bar \theta}_{\mathrm{BV}},r)$ denote the number the LH and RH wave data in the given bins.

The occurrence rate distributions in Figures \ref{fig:sampling}(d)$-$(f) have almost the same trends, that is, they normally decease with increasing ${\bar \theta}_{\mathrm{BV}}$. This indicates that observed ion-scale waves preferentially emerge at smaller ${\bar \theta}_{\mathrm{BV}}$, being consistent with the predictions of the sampling effect \citep{2020ApJS..246...66B}.  Moreover, the occurrence raters in the two near-Sun solar wind regimes ($r=0.05-0.2$ au and $0.2-0.35$ au) are significantly larger than that in the remote solar wind with $r=0.35-0.7$ au at ${\bar \theta}_{\mathrm{BV}}\lesssim 42.5^\circ$. The competition between the occurrence rates in $r=0.05-0.2$ au and $0.2-0.35$ au regimes is dependent on the polarization (LH or RH) of observed waves. For LH waves, the occurrence rate in $r=0.05-0.2$ au is about $1-1.5$ times of that in $r=0.2-0.35$ au at ${\bar \theta}_{\mathrm{BV}}\lesssim 42.5^\circ$; and the former approximates $0.7-1$ times of the latter for RH waves. At large ${\bar \theta}_{\mathrm{BV}}$ (e.g., ${\bar \theta}_{\mathrm{BV}} \gtrsim 42.5^\circ$), the occurrence rate in $r=0.35-0.7$ au is larger than that in $r=0.05-0.2$ au and $0.2-0.35$ au regimes.


Figures \ref{fig:sampling}(g)$-$(i) present the distributions of $N_{\mathrm{wave}}$ $/N_{\mathrm{NOwave}}$, $N_{\mathrm{LH}}/N_{\mathrm{NOwave}}$, and $N_{\mathrm{RH}}/N_{\mathrm{NOwave}}$, respectively. Figures \ref{fig:sampling}(g)$-$(i) exhibit that   $N_{\mathrm{wave}}/N_{\mathrm{NOwave}}$, $N_{\mathrm{LH}}/N_{\mathrm{NOwave}}$, and $N_{\mathrm{RH}}$ $/N_{\mathrm{NOwave}}$ generally decrease as ${\bar \theta}_{\mathrm{BV}}$ increases, and their values in the near-Sun solar wind regimes ($r=0.05-0.2$ au and $0.2-0.35$ au) are larger than those in the remote solar wind with $r=0.35-0.7$ au at ${\bar \theta}_{\mathrm{BV}}\lesssim 42.5^\circ$.
These observational results are in accord with the features of the distributions shown in Figures \ref{fig:sampling}(d)$-$(f).

According to the distributions shown in Figures \ref{fig:sampling}(d)$-$(i), we conclude that  quasi-parallel/antiparallel ion-scale waves in the near-Sun solar wind have higher occurrence rate than those in the remote solar wind under the same sample effect at ${\bar \theta}_{\mathrm{BV}}\lesssim 42.5^\circ$.

\section{The wave mode nature}

Up to now, the wave analysis is based on the spacecraft frame. Now we proceed with the analysis of the nature of observed ion-scale waves in the plasma frame. Due to the Doppler shift effect, RH (or LH) waves observed in the spacecraft frame normally correspond to either the RH (or LH) mode wave propagating outward from the Sun or the LH (or RH) mode wave propagating inward to the Sun in the plasma frame \citep{2020ApJS..246...66B,2020ApJ...890...17Z,2021ApJ...908L..19S}. This ambiguity can be solved by analyzing the correlation between magnetic and velocity fluctuations \citep{2020ApJ...890...17Z}. However, we cannot use this method due to PSP plasma data having the sampling rate lower than the frequency of the waves of interest.

Fortunately, we can conjecture the nature of the observed waves according to recent understanding of the ion composition and the excitation mechanism of ion-scale waves in the near-Sun solar wind \citep{2020ApJS..248....5V,2021ApJ...909....7K,2021ApJ...920..158L,2022ApJ...926..185O}. 
\cite{2020ApJS..248....5V} have shown that the proton beam component can have an unexpected large number density relative to the proton core component, and they have found that most of coexistent ion-scale waves are excited by proton beams. \cite{2021ApJ...909....7K} have studied the linear stability of ion-scale waves under the one- and two-proton component fitting of the proton velocity distribution functions observed by PSP, and they concluded that the wave characteristics are more consistent with the predictions of the two-component fitting model. 
According to these results, we simply assume observed ion-scale waves in the near-Sun solar wind are mainly triggered by free energy associated with the relative drift between the proton beam and core components. We note that based on 309 randomly selected spectra from the Wind observations, \cite{2018PhRvL.120t5102K} have demonstrated that most spectra are unstable in the presence of the proton beam, which implies that the close connection between ion-scale waves and ion beam may hold even in the solar wind near 1 au.

The waves generated by the electromagnetic ion beam instability propagate in the same direction as the ion beam in the plasma frame \citep{2021ApJ...920..158L}. Assuming that this wave excitation mechanism is responsible for the observed waves, a definitive wave propagation direction can be determined, that is, they propagate outward from the Sun in the plasma frame due to the ion beam streaming away from the Sun. Thus, the LH and RH waves observed in the spacecraft frame correspond to LH polarized Alfv\'en ion cyclotron mode and RH polarized fast-magnetosonic whistler mode waves, respectively, as shown by \cite{2020ApJ...890...17Z}.
This inference of the wave propagation direction has been already proposed by \cite{2020ApJ...899...74B}, who suggested that the vast majority of ion-scale waves they observed are antisunward propagation through the analysis of effective phase speeds.

Under the consideration of the wave excitation mechanism being the ion beam instability, one can estimate the wavenumber of the observed waves as well as their frequencies in the plasma frame through the theoretical predictions stated below.

(1) We simply assume that the observed LH polarized Alfv\'en ion cyclotron and RH polarized fast-magnetosonic whistler mode waves can be described by the cold plasma model which contains only protons and electrons. The waves are further assumed to propagate parallel and antiparallel to ${\bm B_0}$. Hence, the dispersion relations of the parallel waves in the plasma frame are approximately given by \citep{2020ApJ...890...17Z},
 \begin{eqnarray}
2\pi f_{\mathrm{pl}}=  \left[ \left(1+\frac{\lambda_p^2k^2}{4}\right)^{1/2} \pm \frac{\lambda_pk}{2} \right] V_Ak,
\label{eq:parallel_pl}
\end{eqnarray}
where $\lambda_p$ denotes the proton inertial length, and the signs ``$+$'' and ``$-$'' in the square bracket represent parallel-propagating fast-mode and Alfv\'en-mode waves, respectively.
We let $f_{\mathrm{pl}}>0$ and $k>0$ in Equation (\ref{eq:parallel_pl}), and this makes the phase velocity $v_p=2\pi f_{\mathrm{pl}}/k>0$ which corresponds to the waves propagating along ${\bm B_0}$.
For the antiparallel waves, their dispersion relations are given by \citep{2020ApJ...890...17Z},
 \begin{eqnarray}
2\pi f_{\mathrm{pl}}=  -\left[ \left(1+\frac{\lambda_p^2k^2}{4}\right)^{1/2} \pm \frac{\lambda_pk}{2} \right] V_Ak.
\label{eq:antiparallel_pl}
\end{eqnarray}
Different from Equation (\ref{eq:parallel_pl}), here $f_{\mathrm{pl}}>0$ and $k<0$, which result in the phase velocity $v_p=2\pi f_{\mathrm{pl}}/k<0$, corresponding to the waves propagating against ${\bm B_0}$. Moreover, the signs ``$+$'' and ``$-$'' in the square bracket represent antiparallel-propagating Alfv\'en-mode and fast-mode waves, respectively.

(2) We then consider the Doppler shift frequency part ${\bm V_0}\cdot{\bm k}$. $V_0>0$ ($V_0<0$) is defined to represent the situation of the solar wind flowing along (against) ${\bm B_0}$. Because we assume the waves being generated by the ion beam instability, they propagate outward from the Sun. Thus, in the case of $V_0>0$ ($V_0<0$), $k>0$ ($k<0$).

(3) By combining $2\pi f_{\mathrm{pl}}$ and ${\bm V_0}\cdot{\bm k}$ parts, we obtain the dispersion relations of the Alfv\'en-mode and fast-mode waves in the spacecraft frame \citep{2020ApJ...890...17Z}, as shown by following equation, 
\begin{eqnarray}
\frac{f_{\mathrm{sc}}}{f_{\mathrm{cp}}} = \frac{\lambda_p {\bm V_0} \cdot {\bm k} }{V_A} \pm \left[ \left(1+\frac{\lambda_p^2k^2}{4}\right)^{1/2} \pm \frac{\lambda_pk}{2} \right] \lambda_pk,
\label{eq:parallel_sc}
\end{eqnarray}
where ``$+$'' (``$-$'') in the first ``$\pm$'' in the right-hand side of the equation denotes the waves in the case of $V_0>0$ ($V_0<0$). 
Using this equation has the advantage of limiting $f_{\mathrm{sc}}$ to a positive value (the wave direction is determined by the sign of $k$). $k$ of observed waves can be directly obtained by inputting the observed $f_{\mathrm{sc}}/f_{\mathrm{cp}}$ and $V_0/V_A$ into Equation (\ref{eq:parallel_sc}).
Once $k$ is determined, the wave frequency $f_{\mathrm{pl}}$ in the plasma frame can be given by

\begin{equation}
f_{\mathrm{pl}} = f_{\mathrm{sc}} - \frac{ {\bm V_0} \cdot {\bm k} } {2\pi}.
\end{equation}

Moreover, according to the distribution of the waves as a function ${\bar \theta}_{\mathrm{BV}}$ shown in Figure 3, we limit the wave data to the range of ${\bar \theta}_{\mathrm{BV}}<30$ (i.e., $\theta_{\mathrm{BV}}<30^\circ$ and $\theta_{\mathrm{BV}}>150^\circ$). Additionally, the criterion of $|n_{\mathrm{SPAN-I}} - n_{\mathrm{SPC}}|$ $/n_{\mathrm{SPAN-I}}<0.5$ is used to select confident plasma measurements in the near-Sun solar wind, where $n_{\mathrm{SPC}}$ and $n_{\mathrm{SPAN-I}}$ denote the number density detected by SPC and SPAN-I instruments, respectively. We collect a total of 72,181 wave data, and nearly all data (71,072 data, being consisting of 53,177 LH wave data and 17,895 RH wave data) reside at $r =0.1-0.3$ au (the wave mode analysis is therefore limited to $r=0.1-0.3$ au). Moreover, the Doppler shift frequency is calculated through ${\bm V_0} \cdot {\bm k}=V_0k\mathrm{cos}(\theta_{\mathrm{kV}})$, where the angle between the solar wind and wavevector $\theta_{\mathrm{kV}}$ equals $\theta_{\mathrm{kV}}=\theta_{\mathrm{BV}}$ ($180-\theta_{\mathrm{BV}}$) in the case of $V_0>0$ ($V_0<0$) under the assumption of the parallel/antiparallel propagation.

\begin{figure}
\includegraphics[width=8.5 cm]{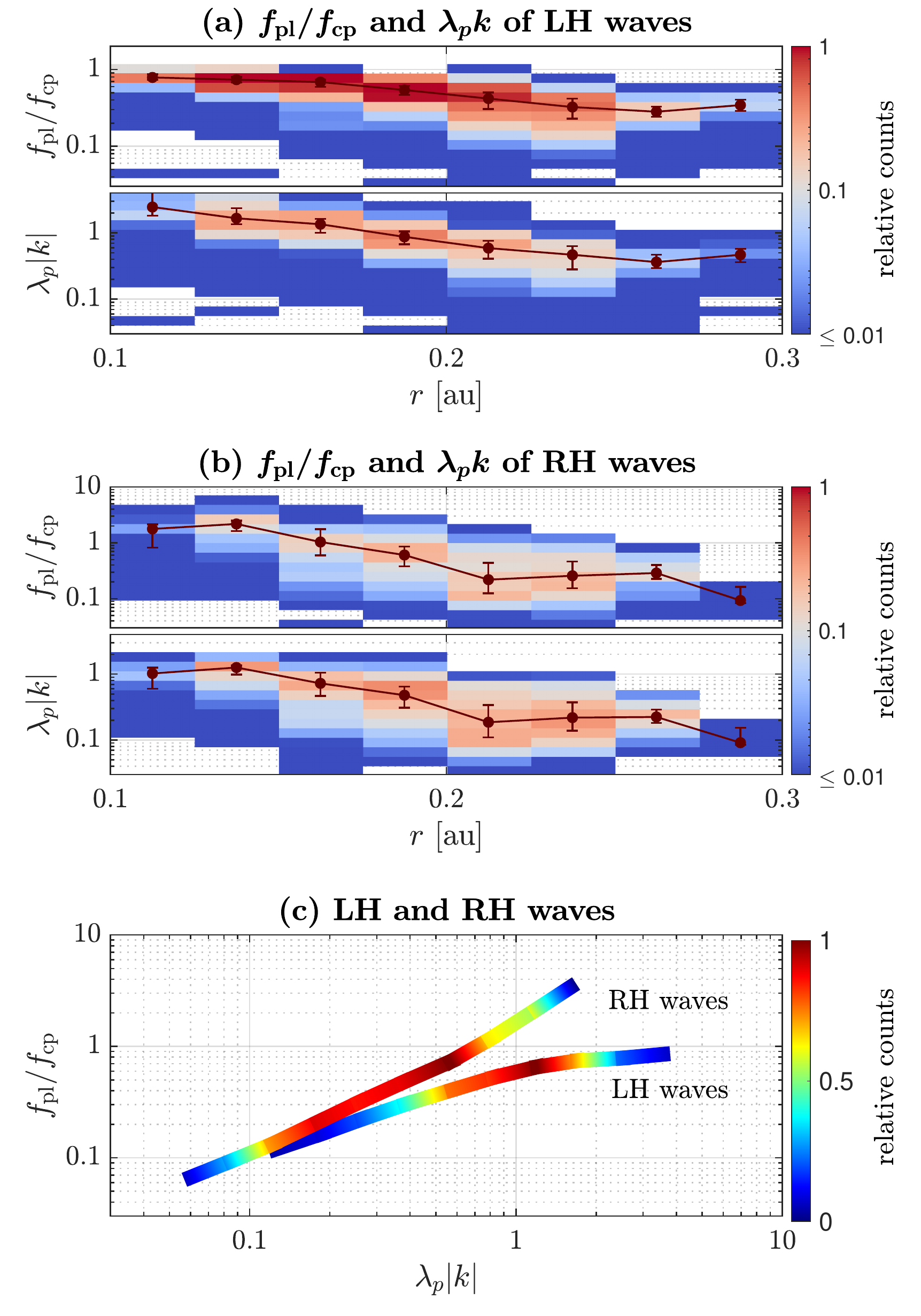}
\caption{
The wave frequency $f_{\mathrm{pl}}$ and wavenumber $k$ in the plasma frame. (a) The radial distribution of $f_{\mathrm{pl}}/f_{\mathrm{cp}}$ and $\lambda_p|k|$ of LH waves; (b) the radial distribution of $f_{\mathrm{pl}}/f_{\mathrm{cp}}$ and $\lambda_p|k|$ of RH waves; and (c) the joint distribution of $f_{\mathrm{pl}}/f_{\mathrm{cp}}$ and $\lambda_p|k|$ for LH and RH waves. The counts in the bins are normalized by the maximum count. The descriptions of the filled points and error bars in Panels (a) and (b) are the same as those in Figure \ref{fig:radial_distribution}.
\label{fig:dispersion_relations}}
\end{figure}

Figures \ref{fig:dispersion_relations}(a) and (b) present the distributions of $f_{\mathrm{pl}}/f_{\mathrm{cp}}$ and $\lambda_pk$ of LH waves and RH waves, respectively.
Figure \ref{fig:dispersion_relations}(a) shows that both $f_{\mathrm{pl}}/f_{\mathrm{cp}}$ and $\lambda_p|k|$ are roughly decreasing with increasing $r$. For RH waves shown in Figure \ref{fig:dispersion_relations}(b), their $f_{\mathrm{pl}}/f_{\mathrm{cp}}$ and $\lambda_p|k|$ at radial distances close to the Sun are also  normally larger than those at remote radial distances. 
Figure \ref{fig:dispersion_relations}(c) presents the joint distributions of $f_{\mathrm{pl}}/f_{\mathrm{cp}}$ and $\lambda_p|k|$ of RH waves (fast-mode waves, upper branch) and LH waves (Alfv\'en-mode waves, lower branch). This figures shows that most (about 76\%) LH wave events concentrate in the range of $f_{\mathrm{pl}}/f_{\mathrm{cp}}\sim 0.3-0.8$ and $\lambda_p|k|\sim 0.3-2$, and about 82\% of RH waves are distributed in the range of $f_{\mathrm{pl}}/f_{\mathrm{cp}}\sim 0.1-1$ and $\lambda_p|k|\sim 0.1-2$.

Although the observed waves are assumed to be driven by free energy carried by the ion beam, it is still unknown whether these waves are produced locally or nonlocally. The answer to this problem requires the instability analysis under the exact information of the ion and electron parameters (e.g., the drift speed and the parallel and perpendicular temperatures of proton core and beam components). However, such an analysis cannot be performed due to lacking of these plasma information. Here, we propose a qualitative estimation for the percentage of the events that are possibly locally excited.
According to the ion beam instability analysis based on fitted plasma parameters in the near-Sun solar wind \citep{2020ApJS..248....5V}, local excitation is nearly inhibited at $\lambda_pk\sim 0.8$, corresponding to $f_{\mathrm{pl}}/f_{\mathrm{cp}}\simeq 0.5$ for LH waves and $f_{\mathrm{pl}}/f_{\mathrm{cp}}\simeq 1$ for RH waves. Using above frequency and wavenumber information, we find that about 40\% (44\%) of LH wave events are limited to $f_{\mathrm{pl}}/f_{\mathrm{cp}} <0.5$ ($\lambda_pk < 0.8$) and about 74\% (85\%) of RH wave events are limited to $f_{\mathrm{pl}}/f_{\mathrm{cp}} <1$ ($\lambda_pk < 0.8$) in data set in Figure \ref{fig:dispersion_relations}. We note that this estimation is sensitive to the frequency and wavenumber thresholds, which are closely related to the plasma parameters \citep{2021ApJ...920..158L}.

\section{Discussions}

\subsection{The sampling effect}

For the wave and turbulence in the solar wind, their observations from a single spacecraft are significantly affected by the sampling effect \citep[e.g.,][]{1976JGR....81.5591F,2010ApJ...709L..49H,2012SSRv..172..325H,2020ApJS..246...66B,2021ApJ...912..101W}. 
Consequently, this sampling effect is one key factor determining the occurrence rate of the observed ion-scale waves \citep{2020ApJS..246...66B}.
The wave occurrence rate distributions shown in Figure \ref{fig:sampling} clearly exhibit the signature of the sampling effect in our wave dataset, that is, $N_{\mathrm{wave}}({\bar \theta}_{\mathrm{BV}},r)/N_{\mathrm{all}}({\bar \theta}_{\mathrm{BV}},r)$ decreases with increasing ${\bar \theta}_{\mathrm{BV}}$ (the angle between the sample direction and the background magnetic field) in different radial regimes. 
Considering the Parker spiral, the sampling effect is one reason inducing the decrease of the wave occurrence rate with increasing radial distance. However, since the wave occurrence rates in the radial regime close to the Sun are larger than those at remote radial distances at small ${\bar \theta}_{\mathrm{BV}}$ (see Figure \ref{fig:sampling}(d)-(f)), we propose that in addition to the sampling effect, there exist other mechanisms leading to the enhancement of ion-scale waves in the near-Sun solar wind. One possible mechanism is that the enhanced ion beam component therein can provide more free energy to emit more stronger ion-scale waves \citep[e.g.,][]{2020ApJS..248....5V,2021ApJ...909....7K,2021ApJ...920..158L}. The checking of this explanation requires the information of the radial distribution of the ion beam population, which will be performed in the future study.

\subsection{The wave identification criteria}

\begin{figure}[t]
\includegraphics[width=8.5 cm]{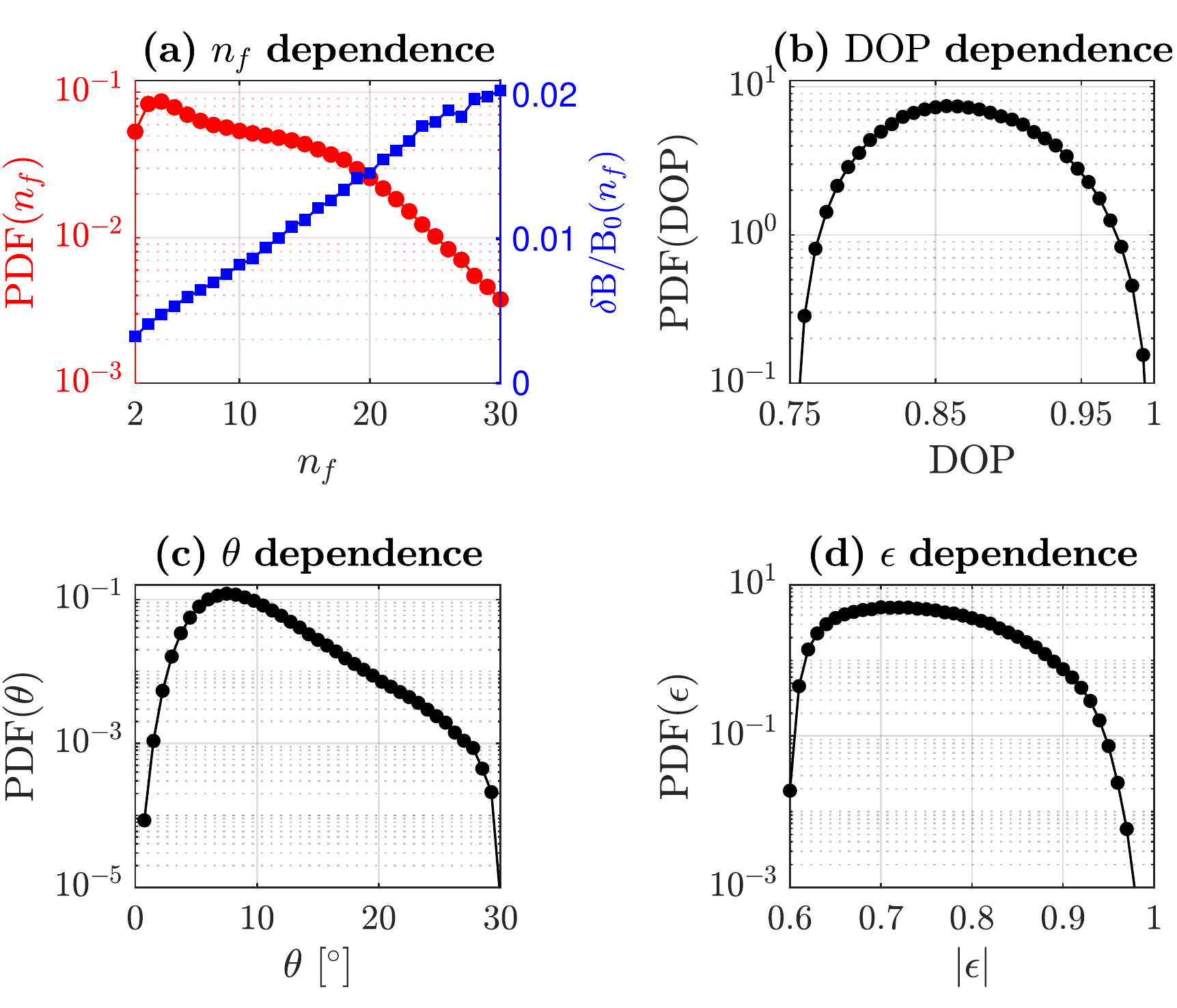}
\caption{
The PDF distributions of the variable: (a) $n_f$, (b) DOP, (c) $\theta$, and (d) $\epsilon$. In panel (a), the blue squares denote the relative amplitude $\delta B/B_0$ at each $n_f$.
\label{fig:parameter_dependence}}
\end{figure}

\begin{figure}
\includegraphics[width=8.5 cm]{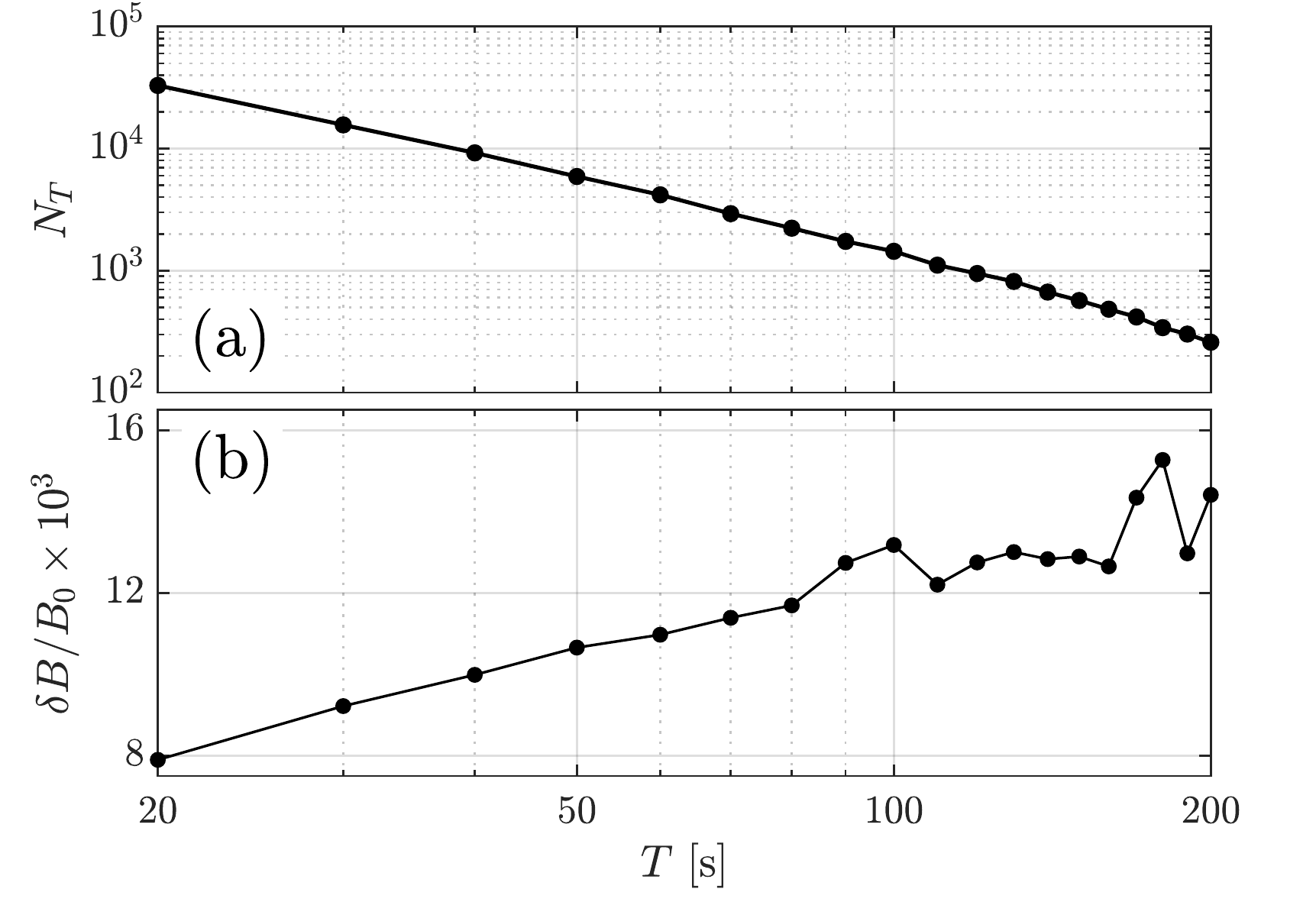}
\caption{
(a) The number and (b) the relative wave amplitude $\delta B/B_0$ as a function of the duration $T$ of the wave event. 
\label{fig:dTdependence}}
\end{figure}

The identification of the wave event is determined by the criteria used in this paper. In order to show that the set of observed waves is robust to the selection criteria, Figure \ref{fig:parameter_dependence} presents the distributions of PDF defined by $\mathrm{PDF}(x) = N(x)/(\Delta x N_t)$ as a function of $x=n_{f}$, $\mathrm{DOP}$, $\theta$, or $\epsilon$ in the wave dataset below 0.3 au. Here, $N(x)$ denotes the number of the wave data in the bins of the variable $x$, $N_t$ denotes the total number of the wave data, and $\Delta x$ is the bin width of the variable $x$.

Figure \ref{fig:parameter_dependence}(a) shows that $\mathrm{PDF}(n_f)$ first increases and then declines as $n_f$ increases. This figure also shows the median value of the relative wave amplitude $\delta B/B_0$ at each $n_f$, which indicates that larger $n_f$ normally corresponds to larger $\delta B/B_0$. When we choose another value of $n_f$ (e.g., $n_f=2$, 3, and $5-10$) instead of $n_f=4$ to select the wave event, the occurrence rate (the ratio between the wave duration and the total observational time) varies as 29\% ($n_f=2$), 28\% ($n_f=3$), 25\% ($n_f=5$), 23\% ($n_f=6$), 21\% ($n_f=7$), 20\% ($n_f=8$), 18\% ($n_f=9$), and 16\% ($n_f=10$) below $r=0.3$ au. These values are much larger than corresponding occurrence rates between 0.3 and 0.7 au, i.e., 13.1\% ($n_f=2$), 12.0\% ($n_f=3$), 8.8\% ($n_f=5$), 7.6\% ($n_f=6$), 6.6\% ($n_f=7$), 5.8\% ($n_f=8$), 5.2\% ($n_f=9$), and 4.6\% ($n_f=10$). Hence, the change of $n_f$ do not alter our conclusion of the enhancement of the occurrence rate of ion-scale waves in the near-Sun solar wind. The choice of $n_f$ does change the occurrence rate of identified waves. However, for smaller $n_f$, the waves have  smaller relative amplitude, and their wave features may be more easily affected by the background turbulence. It should be noted that the criterion of $n_f=4$ used in our wave identification procedure is chosen by the experience, not based on the physical judgment. It needs to develop a physical criterion instead of the  preliminary criterion of $n_f=4$ to pick the waves.

Figures \ref{fig:parameter_dependence}(b)-(d) show that $\mathrm{PDF}$ reaches the maximum at $\mathrm{DOP}\simeq 0.86$, $\theta\simeq 8^\circ$, and $|\epsilon|\simeq 0.71$. Because 99\% of events concentrate at $\mathrm{DOP}\gtrsim 0.77$, $\theta\lesssim 22^\circ$, and $|\epsilon| \gtrsim 0.63$, the dataset of observed waves can be nearly the same as we change the criteria as $\mathrm{DOP} = 0.77$, $\theta = 22^\circ$, and $|\epsilon| = 0.63$.

Figure \ref{fig:dTdependence} further shows the distributions of $N_T$ and the relative wave amplitude $\delta B/B_0$ as a function of the duration $T$ of the whole wave event, where $N_T$ denotes the number of wave data at each $T$. Figure \ref{fig:dTdependence}(a) exhibits $N_T$ increasing with decreasing $T$.
$\delta B/B_0$ shown in Figure \ref{fig:dTdependence}(b) is more larger for the wave event with longer duration. One may be interesting in the information of the wave events with $T<20$ s, according to the change tendency of $\delta B/B_0$ in Figure \ref{fig:dTdependence}(b), these events would have weak amplitudes being about $\delta B/B_0\sim 10^{-3}$. When the $T<20$ s waves are included in our dataset, the wave occurrence rates would be larger than the values explored in this work. The statistics of ion-scale waves below 0.3 au by \cite{2020ApJS..246...66B} explored that the wave number $N_T$ decreases with increasing $T$ as $T$ is smaller than the mean wave duration $\sim 21$ seconds. The measured distribution $N_T(T)$ in the work of \cite{2020ApJS..246...66B} seems to follow the lognomal distribution, indicating that the waves with $T\geq20$ seconds may be the major contributor of the total duration of the waves with no limit of $T$.

\subsection{The wave theory under the actual plasma parameters}

This study uses the wave theory based on cold proton-electron plasmas to estimate the wave frequency and wavenumber in the plasma frame. Because the effects associating with the plasma thermal pressure and alpha particle population play important roles in determining the wave dispersion relation \citep[e.g.,][]{2021ApJ...920..158L,2022ApJ...930...95Z}, our theoretical predictions merely provide preliminary information of the wave parameters. In order to obtain exact wave parameters, one should use the wave theory under the actual plasma parameters, which require accurate fitting of the particle velocity distribution functions measured by PSP. This work will be performed in the future.

\section{Summary}

This work performs a statistical study of ion-scale waves in the inner heliosphere based on PSP observations. Using the SVD method, the waves are selected under the criteria of $\mathrm{DOP}>0.75$, $|\epsilon|>0.6$, and $\theta<30^\circ$. However, due to our limits on the wave duration, i.e., $T\geq 20$ s, this study merely provides a preliminary analysis of the  radial distribution of the waves.

This study finds that the wave frequency and wave amplitude decrease considerably with radial distance below 0.4 au. However, the normalized wave frequency (by the local proton cyclotron frequency) and the normalized wave amplitude (by the local background magnetic field) remain nearly unchanged with radial distance. The wave normal angle and ellipticity also remain nearly unchanged with radial distance. 

The most important finding is that the occurrence rate of ion-scale waves is considerably enhanced in the near-Sun solar wind below 0.3 au. The occurrence rate of LH waves is nearly $1.5-5.8$ times larger than that of RH waves. Also, the occurrence rate in the radial regime close to the Sun is normally higher than that in the remote radial regime as the angle between the sampling direction and the background magnetic field is approximately smaller than $42.5^\circ$. Although the sampling effect induced by one satellite measurements can lead to the decrease of the wave occurrence rate with radial distance, the wave enhancement in the near-Sun solar wind can also result from local wave excitation mechanisms. For example, the ion beam instability has recently been proposed as a source of ion-scale waves in the near-Sun solar wind \citep{2020ApJS..248....5V,2021ApJ...920..158L}. 

For the mode nature of the observed ion-scale waves, this work discriminates their characteristics based on the assumption that they are driven by the ion beam instability. We propose that LH and RH waves in the spacecraft frame correspond to LH polarized Alfv\'en ion cyclotron waves and RH polarized fast-magnetosonic whistler wave in the plasma frame. By using the wave theory in the cold plasma model, we estimated the wavenumber $k$ of the observed waves and their frequencies $f_{\mathrm{pl}}$ in the plasma frame, and we found that both $\lambda_p k$ and $f_{\mathrm{pl}}/f_{\mathrm{cp}}$ are roughly decreasing with radial distance below 0.3 au. 

These findings can provide useful insights into the evolution of ion-scale waves and their role in particle dynamics in the near-Sun solar wind.

\begin{acknowledgments}
This work was supported by the National Key R\&D Program of China 2021YFA1600502 (2021YFA1600500) and the NSFC 41974203. We appreciate the referee for helpful suggestions and inspiring comments. Z.J.S. also appreciates Dr. Jia Huang for useful discussions.
Parker Solar Probe was designed, built, and is now operated by the Johns Hopkins Applied Physics Laboratory as part of NASA’s Living with a Star (LWS) program (contract NNN06AA01C). All data used in this paper are publicly available from the links http://fields.ssl.berkeley.edu/data/ 
and http://sweap. cfa.harvard.edu/pub/data/sci/sweap/. The wave analyses were performed by using the IRFU-Matlab package (https://github.com/irfu/irfu-matlab). 
L.W. performed the statistics, and contributed to the data analysis and the preparation of the manuscript. 
Z.J.S. initiated the study, analyzed the data, and prepared the manuscript.
W.T.Y and D.X.C. contributed to the development of the wave identification procedure. 
J.C.K is the PI of the SWEAP, and S.D.B is the PI of the FIELDS. 
S.C. and W.D.J contributed to in the wave analysis.

\end{acknowledgments}



\clearpage

\end{document}